\documentclass[twocolumn]{article}
\usepackage{latexsym}
\usepackage{graphics}

\def\comment #1{}

\def\refer #1{{(\ref{#1})}}

\def\of #1{\!\left({#1}\right)}

\def\set #1{\left\lbrace{#1}\right\rbrace}

\def\brackets #1{\left[{#1}\right]}

\def\commutator #1#2{\brackets{{#1},{#2}}}

\def\defas {:=}

\def\eigenspace #1#2 {\mathrm{eig}\of{#1,#2}}

\newlength{\skiplength}

\def\inner {\!\cdot\!}

\def\@spires#1{\href{http://www-spires.slac.stanford.edu/spires/find/hep/www?j=#1}} 
\catcode`\%=12
\catcode`\|=14
\newcommand\apa[3]    {\@spires{APASA%2C#1%2C#3}
		{{\it Acta Phys.\ Austriaca }{\bf #1} (#2) #3}}
\newcommand\apas[3]    {\@spires{APAUA%2C#1%2C#3}
		{{\it Acta Phys.\ Austriaca, Suppl.\ }{\bf #1} (#2) #3}}
\newcommand\appol[3] {\@spires{APPOA%2C#1%2C#3}
		{{\it Acta Phys.\ Polon.\ }{\bf #1} (#2) #3}}
\newcommand\advm[3]  {\@spires{ADMTA%2C#1%2C#3}
		{{\it Adv.\ Math.\ }{\bf #1} (#2) #3}}
\newcommand\adnp[3]   {\@spires{ANUPB%2C#1%2C#3}
		{{\it Adv.\ Nucl.\ Phys.\ }{\bf #1} (#2) #3}}
\newcommand\adp[3]   {\@spires{ADPHA%2C#1%2C#3}
		{{\it Adv.\ Phys.\ }{\bf #1} (#2) #3}}
\newcommand\atmp[3] {\@spires{00203%2C#1%2C#3}
		{{\it Adv.\ Theor.\ Math.\ Phys.\ }{\bf #1} (#2) #3}}
\newcommand\am[3]    {\@spires{ANMAA%2C#1%2C#3}
		{{\it Ann.\ Math.\ }{\bf #1} (#2) #3}}
\newcommand\ap[3]    {\@spires{APNYA%2C#1%2C#3}
		{{\it Ann.\ Phys.\ (NY) }{\bf #1} (#2) #3}}
\newcommand\araa[3] {\@spires{ARAAA%2C#1%2C#3}
		{{\it Ann.\ Rev.\ Astron.\ \& Astrophys.\ }{\bf #1} (#2) #3}}
\newcommand\arnps[3] {\@spires{ARNUA%2C#1%2C#3}
		{{\it Ann.\ Rev.\ Nucl.\ Part.\ Sci.\ }{\bf #1} (#2) #3}}
\newcommand\asas[3]   {\@spires{AAEJA%2C#1%2C#3}
		{{\it Astron.\ Astrophys.\ }{\bf #1} (#2) #3}}
\newcommand\asj[3]   {\@spires{ANJOA%2C#1%2C#3}
		{{\it Astron.\ J.\ }{\bf #1} (#2) #3}}
\newcommand\app[3]   {\@spires{APHYE%2C#1%2C#3}
		{{\it Astropart.\ Phys.\ }{\bf #1} (#2) #3}}
\newcommand\apj[3]    {\@spires{ASJOA%2C#1%2C#3}
		{{\it Astrophys.\ J. }{\bf #1} (#2) #3}}
\newcommand\baas[3]   {\@spires{AASBA%2C#1%2C#3}
		{{\it Bull.\ Am.\ Astron.\ Soc.\ }{\bf #1} (#2) #3}}
\newcommand\bams[3]   {\@spires{BAMOA%2C#1%2C#3}
		{{\it Bull.\ Am.\ Math.\ Soc.\ }{\bf #1} (#2) #3}}
\newcommand\blms[3]   {\@spires{LMSBB%2C#1%2C#3}
		{{\it Bull.\ London Math.\ Soc.\ }{\bf #1} (#2) #3}}
\newcommand\cjm[3]  {\@spires{CJMAA%2C#1%2C#3}
		{{\it Can.\ J.\ Math.\ }{\bf #1} (#2) #3}}
\newcommand\cqg[3]   {\@spires{CQGRD%2C#1%2C#3}
		{{\it Class.\ and Quant.\ Grav.\ }{\bf #1} (#2) #3}}
\newcommand\cmp[3]   {\@spires{CMPHA%2C#1%2C#3}
		{{\it Commun.\ Math.\ Phys.\ }{\bf #1} (#2) #3}}
\newcommand\ctp[3]   {\@spires{CTPMD%2C#1%2C#3}
		{{\it Commun.\ Theor.\ Phys.\ }{\bf #1} (#2) #3}}
\newcommand\cag[3]   {\@spires{00142%2C#1%2C#3}
		{{\it Commun.\ Anal.\ Geom.\ }{\bf #1} (#2) #3}}
\newcommand\cpam[3]   {\@spires{CPAMA%2C#1%2C#3}
		{{\it Commun.\ Pure Appl.\ Math.\ }{\bf #1} (#2) #3}}
\newcommand\cpc[3]   {\@spires{CPHCB%2C#1%2C#3}
		{{\it Comput.\ Phys.\ Commun.\ }{\bf #1} (#2) #3}}
\newcommand\dmj[3]   {\@spires{DUMJA%2C#1%2C#3}
		{{\it Duke Math.\ J. }{\bf #1} (#2) #3}}
\newcommand\epjc[3]  {\@spires{EPHJA%2CC#1%2C#3}
		{{\it Eur.\ Phys.\ J. }{\bf C #1} (#2) #3}}
\newcommand\epjd[3]  {\@spires{EPHJD%2CC#1%2C#3}
		{{\it Eur.\ Phys.\ J. Direct.\ }{\bf C #1} (#2) #3}}
\newcommand\epl[3]    {\@spires{EULEE%2C#1%2C#3}
		{{\it Europhys.\ Lett. }{\bf #1} (#2) #3}}
\newcommand\forp[3]    {\@spires{FPYKA%2C#1%2C#3}
		{{\it Fortschr.\ Phys.\ }{\bf #1} (#2) #3}}
\newcommand\faa[3]    {\@spires{FAAPB%2C#1%2C#3}
		{{\it Funct.\ Anal.\ Appl.\ }{\bf #1} (#2) #3}}
\newcommand\grg[3]    {\@spires{GRGVA%2C#1%2C#3}
		{{\it Gen.\ Rel.\ Grav.\ }{\bf #1} (#2) #3}}
\newcommand\hpa[3]   {\@spires{HPACA%2C#1%2C#3}
		{{\it Helv.\ Phys.\ Acta }{\bf #1} (#2) #3}}
\newcommand\ijmpa[3] {\@spires{IMPAE%2CA#1%2C#3}
		{{\it Int.\ J.\ Mod.\ Phys.\ }{\bf A #1} (#2) #3}}
\newcommand\ijmpb[3] {\@spires{IMPAE%2CB#1%2C#3}
		{{\it Int.\ J.\ Mod.\ Phys.\ }{\bf B #1} (#2) #3}}
\newcommand\ijmpc[3] {\@spires{IMPAE%2CC#1%2C#3}
		{{\it Int.\ J.\ Mod.\ Phys.\ }{\bf C #1} (#2) #3}}
\newcommand\ijmpd[3] {\@spires{IMPAE%2CD#1%2C#3}
		{{\it Int.\ J.\ Mod.\ Phys.\ }{\bf D #1} (#2) #3}}
\newcommand\ijtp[3] {\@spires{IJTPB%2CB#1%2C#3}
		{{\it Int.\ J.\ Theor.\ Phys.\ }{\bf #1} (#2) #3}}
\newcommand\invm[3]  {\@spires{INVMB%2C#1%2C#3}
		{{\it Invent.\ Math.\ }{\bf #1} (#2) #3}}
\newcommand\jag[3]   {\@spires{00124%2C#1%2C#3}
		{{\it J.\ Alg.\ Geom.\ }{\bf #1} (#2) #3}}
\newcommand\jams[3]   {\@spires{00052%2C#1%2C#3}
		{{\it J.\ Am.\ Math.\ Soc.\ }{\bf #1} (#2) #3}}
\newcommand\jap[3]   {\@spires{JAPIA%2C#1%2C#3}
		{{\it J.\ Appl.\ Phys.\ }{\bf #1} (#2) #3}}
\newcommand\jdg[3]   {\@spires{JDGEA%2C#1%2C#3}
		{{\it J.\ Diff.\ Geom.\ }{\bf #1} (#2) #3}}
\newcommand\jgp[3]   {\@spires{JGPHE%2C#1%2C#3}
		{{\it J.\ Geom.\ Phys.\ }{\bf #1} (#2) #3}}
\newcommand\jhep[3]  {\href{http://jhep.sissa.it/stdsearch?paper=#1%28#2%29#3}
		{{\it J. High Energy Phys.\ }{\bf #1} (#2) #3}}
\newcommand\jmp[3]   {\@spires{JMAPA%2C#1%2C#3}
		{{\it J.\ Math.\ Phys.\ }{\bf #1} (#2) #3}}
\newcommand\joth[3]  {\@spires{JOTHE%2C#1%2C#3}
		{{\it J.\ Operator Theory }{\bf #1} (#2) #3}}
\newcommand\jpha[3]   {\@spires{JPAGB%2CA#1%2C#3}
		{{\it J. Phys.\ }{\bf A #1} (#2) #3}}
\newcommand\jphc[3]   {\@spires{JPAGB%2CC#1%2C#3}
		{{\it J. Phys.\ }{\bf C #1} (#2) #3}}
\newcommand\jphg[3]   {\@spires{JPAGB%2CG#1%2C#3}
		{{\it J. Phys.\ }{\bf G #1} (#2) #3}}
\newcommand\lmp[3]   {\@spires{LMPHD%2CA#1%2C#3}
		{{\it Lett.\ Math.\ Phys.\ }{\bf #1} (#2) #3}}
\newcommand\ncl[3]    {\@spires{NCLTA%2C#1%2C#3}
		{{\it Lett.\ Nuovo Cim.\ }{\bf #1} (#2) #3}}
\newcommand\matan[3]  {\@spires{MAANA%2CA#1%2C#3}
		{{\it Math.\ Ann.\ }{\bf #1} (#2) #3}}
\newcommand\mussr[3]  {\@spires{MUSIA%2CA#1%2C#3}
		{{\it Math.\ USSR Izv.\ }{\bf #1} (#2) #3}}
\newcommand\mams[3]  {\@spires{MAMCA%2CA#1%2C#3}
		{{\it Mem.\ Am.\ Math.\ Soc.\ }{\bf #1} (#2) #3}}
\newcommand\mpla[3]  {\@spires{MPLAE%2CA#1%2C#3}
		{{\it Mod.\ Phys.\ Lett.\ }{\bf A #1} (#2) #3}}
\newcommand\mplb[3]  {\@spires{MPLAE%2CB#1%2C#3}
		{{\it Mod.\ Phys.\ Lett.\ }{\bf B #1} (#2) #3}}
\newcommand\nature[3]  {\@spires{NATUA%2C#1%2C#3}
		{{\it Nature }{\bf #1} (#2) #3}}
\newcommand\nim[3]   {\@spires{NUIMA%2C#1%2C#3}
		{{\it Nucl.\ Instrum.\ Meth.\ }{\bf #1} (#2) #3}}
\newcommand\npa[3]   {\@spires{NUPHA%2CA#1%2C#3}
		{{\it Nucl.\ Phys.\ }{\bf A #1} (#2) #3}}
\newcommand\npb[3]    {\@spires{NUPHA%2CB#1%2C#3}
		{{\it Nucl.\ Phys.\ }{\bf B #1} (#2) #3}}
\newcommand\npps[3]  {\@spires{NUPHZ%2C#1%2C#3}
		{{\it Nucl.\ Phys.\ }{\bf #1} {\it(Proc.\ Suppl.)} (#2) #3}}
\newcommand\nc[3]    {\@spires{NUCIA%2C#1%2C#3}
		{{\it Nuovo Cim.\ }{\bf #1} (#2) #3}}
\newcommand\ncs[3]  {\@spires{NUCUA%2C#1%2C#3}
		{{\it Nuovo Cim.\ Suppl.\ }{\bf #1} (#2) #3}}
\newcommand\pan[3]  {\@spires{PANUE%2C#1%2C#3}
		{{\it Phys.\ Atom.\ Nucl.\ }{\bf #1} (#2) #3}}
\newcommand\pla[3]   {\@spires{PHLTA%2CA#1%2C#3}
		{{\it Phys.\ Lett.\ }{\bf A #1} (#2) #3}}
\newcommand\plb[3]   {\@spires{PHLTA%2CB#1%2C#3}
		{{\it Phys.\ Lett.\ }{\bf B #1} (#2) #3}}
\newcommand\pr[3]    {\@spires{PHRVA%2C#1%2C#3}
		{{\it Phys.\ Rev.\ }{\bf #1} (#2) #3}}
\newcommand\pra[3]   {\@spires{PHRVA%2CA#1%2C#3}
		{{\it Phys.\ Rev.\ }{\bf A #1} (#2) #3}}
\newcommand\prb[3]   {\@spires{PHRVA%2CB#1%2C#3}
		{{\it Phys.\ Rev.\ }{\bf B #1} (#2) #3}}
\newcommand\prc[3]   {\@spires{PHRVA%2CC#1%2C#3}
		{{\it Phys.\ Rev.\ }{\bf C #1} (#2) #3}}
\newcommand\prd[3]   {\@spires{PHRVA%2CD#1%2C#3}
		{{\it Phys.\ Rev.\ }{\bf D #1} (#2) #3}}
\newcommand\pre[3]   {\@spires{PHRVA%2CE#1%2C#3}
		{{\it Phys.\ Rev.\ }{\bf E #1} (#2) #3}}
\newcommand\prep[3]  {\@spires{PRPLC%2C#1%2C#3}
		{{\it Phys.\ Rept.\ }{\bf #1} (#2) #3}}
\newcommand\prl[3]   {\@spires{PRLTA%2C#1%2C#3}
		{{\it Phys.\ Rev.\ Lett.\ }{\bf #1} (#2) #3}}
\newcommand\phys[3]   {\@spires{PHYSA%2CA#1%2C#3}
		{{\it Physica }{\bf #1} (#2) #3}}
\newcommand\plms[3]   {\@spires{PHLTA%2CB#1%2C#3}
		{{\it Proc.\ London Math.\ Soc.\ }{\bf B #1} (#2) #3}}
\newcommand\pnas[3]  {\@spires{PNASA%2C#1%2C#3}
		{{\it Proc.\ Nat.\ Acad.\ Sci.\ }{\bf #1} (#2) #3}}
\newcommand\ppnp[3]  {\@spires{PPNPD%2C#1%2C#3}
		{{\it Prog.\ Part.\ Nucl.\ Phys.\ }{\bf #1} (#2) #3}}
\newcommand\ptp[3]   {\@spires{PTPKA%2C#1%2C#3}
		{{\it Prog.\ Theor.\ Phys.\ }{\bf #1} (#2) #3}}
\newcommand\ptps[3]   {\@spires{PTPSA%2C#1%2C#3}
		{{\it Prog.\ Theor.\ Phys.\ Suppl.\ }{\bf #1} (#2) #3}}
\newcommand\rmp[3]   {\@spires{RMPHA%2C#1%2C#3}
		{{\it Rev.\ Mod.\ Phys.\ }{\bf #1} (#2) #3}}
\newcommand\sjnp[3]  {\@spires{SJNCA%2C#1%2C#3}
		{{\it Sov.\ J.\ Nucl.\ Phys.\ }{\bf #1} (#2) #3}}
\newcommand\sjpn[3]  {\@spires{SJPNA%2C#1%2C#3}
		{{\it Sov.\ J.\ Part.\ Nucl.\ }{\bf #1} (#2) #3}}
\newcommand\jetp[3]  {\@spires{SPHJA%2C#1%2C#3}
		{{\it Sov.\ Phys.\ JETP\/ }{\bf #1} (#2) #3}}
\newcommand\jetpl[3]  {\@spires{JTPLA%2C#1%2C#3}
		{{\it Sov.\ Phys.\ JETP Lett.\ }{\bf #1} (#2) #3}}
\newcommand\spu[3]  {\@spires{SOPUA%2C#1%2C#3}
		{{\it Sov.\ Phys.\ Usp.\ }{\bf #1} (#2) #3}}
\newcommand\tmf[3]   {\@spires{TMFZA%2C#1%2C#3}
		{{\it Teor.\ Mat.\ Fiz.\ }{\bf #1} (#2) #3}}
\newcommand\tmp[3]   {\@spires{TMPHA%2C#1%2C#3}
		{{\it Theor.\ Math.\ Phys.\ }{\bf #1} (#2) #3}}
\newcommand\ufn[3]   {\@spires{UFNAA%2C#1%2C#3}
		{{\it Usp.\ Fiz.\ Nauk.\ }{\bf #1} (#2) #3}}
| }}}}}}}}}}}}}}}}}}}}}} "|" is here a comment (catcode defined above) to
| }}}}}}}}}}}}}}}}}}}}}} include parenthesis for emacs to parse properly. 
\newcommand\ujp[3]   {\@spires{00267%2C#1%2C#3}
		{{\it Ukr.\ J.\ Phys.\ }{\bf #1} (#2) #3}}
\newcommand\yf[3]    {\@spires{YAFIA%2C#1%2C#3}
		{{\it Yad.\ Fiz.\ }{\bf #1} (#2) #3}}
\newcommand\zpc[3]   {\@spires{ZEPYA%2CC#1%2C#3}
		{{\it Z.\ Physik }{\bf C #1} (#2) #3}}
\newcommand\zetf[3]  {\@spires{ZETFA%2C#1%2C#3}
		{{\it Zh.\ Eksp.\ Teor.\ Fiz.\ }{\bf #1} (#2) #3}}

\newcommand{\newjournal}[5]{\@spires{#2%2C#3%2C#5}
		{{\it #1 }{\bf #3} (#4) #5}}

\newcommand\ibid[3]{{\it ibid.\ }{\bf #1} (#2) #3}
\catcode`\%=14
\catcode`\|=12
\newcommand{\hepth}[1]{\href{http://xxx.lanl.gov/abs/hep-th/#1}{\tt hep-th/#1}}
\newcommand{\hepph}[1]{\href{http://xxx.lanl.gov/abs/hep-ph/#1}{\tt hep-ph/#1}}
\newcommand{\heplat}[1]{\href{http://xxx.lanl.gov/abs/hep-lat/#1}{\tt hep-lat/#1}}
\newcommand{\hepex}[1]{\href{http://xxx.lanl.gov/abs/hep-ex/#1}{\tt hep-ex/#1}}
\newcommand{\nuclth}[1]{\href{http://xxx.lanl.gov/abs/nucl-th/#1}{\tt nucl-th/#1}}
\newcommand{\nuclex}[1]{\href{http://xxx.lanl.gov/abs/nucl-ex/#1}{\tt nucl-ex/#1}}
\newcommand{\grqc}[1]{\href{http://xxx.lanl.gov/abs/gr-qc/#1}{\tt gr-qc/#1}}
\newcommand{\qalg}[1]{\href{http://xxx.lanl.gov/abs/q-alg/#1}{\tt q-alg/#1}}
\newcommand{\accphys}[1]{\href{http://xxx.lanl.gov/abs/accphys/#1}{\tt accphys/#1}}
\newcommand{\alggeom}[1]{\href{http://xxx.lanl.gov/abs/alg-geom/#1}{\tt alg-geom/#1}}
\newcommand{\astroph}[1]{\href{http://xxx.lanl.gov/abs/astro-ph/#1}{\tt astro-ph/#1}}
\newcommand{\chaodyn}[1]{\href{http://xxx.lanl.gov/abs/chao-dyn/#1}{\tt chao-dyn/#1}}
\newcommand{\condmat}[1]{\href{http://xxx.lanl.gov/abs/cond-mat/#1}{\tt cond-mat/#1}}
\newcommand{\nlinsys}[1]{\href{http://xxx.lanl.gov/abs/nlin-sys/#1}{\tt nlin-sys/#1}}
\newcommand{\quantph}[1]{\href{http://xxx.lanl.gov/abs/quant-ph/#1}{\tt quant-ph/#1}}
\newcommand{\solvint}[1]{\href{http://xxx.lanl.gov/abs/solv-int/#1}{\tt solv-int/#1}}
\newcommand{\suprcon}[1]{\href{http://xxx.lanl.gov/abs/supr-con/#1}{\tt supr-con/#1}}
\newcommand{\Math}[2]{\href{http://xxx.lanl.gov/abs/math.#1/#2}{\tt math.#1/#2}}

\def\hepth #1{{\tt hep-th/#1}}
\def\ap #1#2#3{{\rm Ann. Phys.} {\bf #1} {\rm #3} {\rm (#2)}}
\def\cmp #1#2#3{{\rm Commun. Math. Phys.} {\bf #1} {\rm #3} {\rm (#2)}}
\def\ijmpd #1#2#3{{\rm Int. J. Mod. Phys. A} {\bf #1} {\rm #3} {\rm (#2)}}

\bibliographystyle{JHEP}

\begin{document}

\title{Pohlmeyer invariants are expressible in terms of DDF invariants}
\author{Urs Schreiber\thanks{Urs.Schreiber@uni-essen.de} \\ Universit{\"a}t Duisburg-Essen \\ Essen, 45117, Germany}
\maketitle

\emph{
It is shown that the Pohlmeyer invariants of the classical bosonic string
are a proper subset of the classical DDF invariants. This makes the quantization
of the Pohlmeyer invariants particularly transparent and allows to generalize them
to the superstring. 
}

The \emph{Pohlmeyer program} 
\cite{Pohlmeyer:2002,MeusburgerRehren:2002,Pohlmeyer:1998,Pohlmeyer:1988,Thiemann:2004,Bahns:2004}
is an attempt to quantize the bosonic string by finding a constistent quantum
deformation of the Poisson-algebra of a certain set of classical invariants, 
the so-called \emph{Pohlmeyer invariants}. So far this
program has succeeded only up to the as yet unproven \emph{quadratic generation hypothesis}
\cite{MeusburgerRehren:2002}. The hope has been expressed that after completion
the Pohlmeyer program would yield an alternative quantization of the string which
does not feature the usual quantum effects like the critical dimension.

Here we show that the Pohlmeyer invariants can consistently be quantized by expressing them
in terms of DDF invartiants whose quantization as DDF operators is well known and leads
to the standard theory. The result reported in this letter is discussed in 
more detail in \cite{Schreiber:2004b}.

Let $X\of{\sigma}$ and $P\of{\sigma}$ be canonical coordinates and
momenta of the bosonic string with Poisson brackets
\begin{eqnarray}
  \commutator{X^\mu\of{\sigma}}{P_\nu\of{\kappa}}_\mathrm{PB}
  &=&
  \delta^\mu_\nu\, \delta\of{\sigma-\kappa}
  \nonumber
\end{eqnarray}
and define
\begin{eqnarray}
  \mathcal{P}_\pm^\mu(\sigma)
  &=&
  \frac{1}{\sqrt{2T}}\left(
    P^\mu\of{\sigma} \pm T X^{\prime \mu}\of{\sigma}
  \right)
  \,.
  \nonumber
\end{eqnarray}
(Here $X^\prime = \partial_\sigma X$, $T = 1/2\pi \alpha^\prime$ is the string tension and we assume a trivial
Minkowski background and shift all spacetime indices with 
$\eta_{\mu\nu} = \mathrm{diag}(-1,1,\cdots,1)$.)

Using the usual oscillators
\begin{eqnarray}
  \label{oscillator expansion}
  \mathcal{P}^\mu_+\of{\sigma}
  &\defas&
  \frac{1}{\sqrt{2\pi}}\sum_m \tilde \alpha_m^\mu e^{-im\sigma}
  \nonumber\\
  \mathcal{P}^\mu_-\of{\sigma}
  &\defas&
  \frac{1}{\sqrt{2\pi}}\sum_m \alpha_m^\mu e^{+im\sigma}
  \nonumber
\end{eqnarray}
and center of mass coordinates and momenta
\begin{eqnarray}
  x^\mu &\defas& \frac{1}{2\pi}\int X^\mu\of{\sigma}\,d\sigma
  \nonumber\\
  p^\mu &\defas& \int P^\mu\of{\sigma}\, d\sigma \;=\; \frac{1}{\sqrt{4\pi T}}\alpha_0 
    = \frac{1}{\sqrt{4\pi T}}\tilde \alpha_0
  \nonumber
\end{eqnarray}
we can write down the left- and right-moving fields of vanishing conformal weight
\begin{eqnarray}
  X^\mu_-\of{\sigma}
  &\defas&
  x^\mu
  -
  \frac{\sigma}{4\pi T}p^\mu
  +
  \frac{i}{\sqrt{4\pi T}}\sum_{m\neq 0} \frac{1}{m}\alpha_m^\mu e^{+im\sigma}
  \nonumber
\end{eqnarray}
and
\begin{eqnarray}
  X^\mu_+\of{\sigma}
  &\defas&
  x^\mu
  +
  \frac{\sigma}{4\pi T}p^\mu
  +
  \frac{i}{\sqrt{4\pi T}}\sum_{m\neq 0} \frac{1}{m}\tilde \alpha_m^\mu e^{-im\sigma}
  \,.
  \nonumber
\end{eqnarray}
By fixing an arbitrary lightlike vector field $k$ on target space we define the objetcs
\begin{eqnarray}
  R_\pm\of{\sigma}
  &\defas&
  \pm
  \frac{4\pi T}{k\inner p}\,
    k\inner X_\pm\of{\sigma}
  \,.
\end{eqnarray}
which are normalized such that
\begin{eqnarray}
  R_\pm\of{\sigma + 2\pi}
  &=&
  R_\pm\of{\sigma} + 2\pi
  \,.
\end{eqnarray}
Their derivative is
\begin{eqnarray}
  R^\prime_\pm\of{\sigma}
  &=&
  \frac{2\pi\sqrt{2T}}{k\inner p}\, k\inner \mathcal{P}_\pm\of{\sigma}
\end{eqnarray}
and can be seen to be non-vanishing for all $\sigma$ on all of phase space
except for a set of measure 0, as discussed in \cite{Schreiber:2004b}.

Now the classical DDF observables $A_m^\mu$ and $\tilde A_m^\mu$ of the closed bosonic string are defined 
by
\begin{eqnarray}
  \label{definition classical DDF}
  A_m^\mu
  &\defas&
  \frac{1}{\sqrt{2\pi}}
  \int d \sigma\,
  \mathcal{P}_-^\mu\of{\sigma}
  e^{
    -im R_-\of{\sigma}
  }
  \nonumber\\
  \tilde A_m^\mu
  &\defas&
  \frac{1}{\sqrt{2\pi}}
  \int d \sigma\,
  \mathcal{P}_+^\mu\of{\sigma}
  e^{
    im R_+\of{\sigma}
  }
  \,.
\end{eqnarray} 
The coordinate 0-mode $k\inner x$ in $R_\pm$ couples the left- and right-moving
Virasoro algebras. Splitting off this factor yields the `truncated' observables
\begin{eqnarray}
  a^\mu_m &\defas& A^\mu_m e^{-im \frac{2T}{k\inner p}\,k\inner x}
  \nonumber\\
  \tilde a^\mu_m &\defas& A^\mu_m e^{-im \frac{2T}{k\inner p}\,k\inner x}
  \,.
\end{eqnarray}

It is now easy to see that the objects
\begin{eqnarray}
  \label{DDF invariants}
  D_{\set{m_i,\tilde m_j}} 
  &\defas& 
  a^{\mu_1}_{m_1}\cdots a^{\mu_r}_{m_r}\,\tilde a^{\nu_1}_{\tilde n_1}\cdots a^{\nu_s}_{\tilde m_s}
  e^{i N \frac{2T}{k\inner p}\, k\inner x}
\end{eqnarray}
when satisfying the \emph{level matching condition}
\begin{eqnarray}
  \label{level matching condition}
  \sum\limits_i m_i \;=\; N \;=\;  \sum\limits_j \tilde m_j
  \,,
\end{eqnarray}
Poisson-commute with all the Virasoro constraints, i.e. with $(\mathcal{P}_\pm)^2\of{\sigma}\,,\forall \sigma$.
This are the \emph{classical DDF invariants} of the classical closed bosonic string
in flat target space.\\

From the Fourier mode-like objects $A_m^\mu$ and $\tilde A_m^\mu$ one reobtains
quasi-local fields
$\mathcal{P}^R_\pm$ by an inverse Fourier transformation:
\begin{eqnarray}
  \mathcal{P}^R_-\of{\sigma}
  &\defas&
  \frac{1}{\sqrt{2\pi}}
  \sum\limits_m A_m^\mu e^{+im\sigma}
  \nonumber\\
  \mathcal{P}^R_+\of{\sigma}
  &\defas&
  \frac{1}{\sqrt{2\pi}}
  \sum\limits_m \tilde A_m^\mu e^{-im\sigma}
  \,.
\end{eqnarray}
Since $R_\pm$ is invertible almost everywhere on phase space this is equal to
\begin{eqnarray}
  \mathcal{P}^R_\pm\of{\sigma}
  &=&
  \left((R_\pm)^{-1}\right)^\prime\of{\sigma}
  \mathcal{P}^\mu\of{(R_\pm)^{-1}\of{\sigma}}
  \,.
\end{eqnarray}
But this is just the formula for the transformation of the
unit weight object $\mathcal{P}_\pm$ under the reparameterization 
$\sigma \mapsto R^{-1}_\pm\of{\sigma}$. This means that any functional 
$F\of{\mathcal{P}_\pm}$ of the
original $\mathcal{P}_\pm$ which is reparameterization invariant remains
invariant after substition of $\mathcal{P}^R_\pm$ for $\mathcal{P}_\pm$:
\begin{eqnarray}
  \label{reparameterization property}
  F\of{\mathcal{P}_\pm} &=& F\of{\mathcal{P}^R_\pm}
  \,.
\end{eqnarray}
This way all such
reparameterization invariant functionals can be re-expressed in terms of
DDF invariants.\\

One particular reparameterization invariant functional is the \emph{Wilson line}.
Let $A_\mu$ be any \emph{constant} gauge connection on target space, then
\begin{eqnarray}
  F\of{\mathcal{P}_\pm}
  &\defas&
  \mathrm{Tr}\mathbf{P}
  \exp\of{ \int\limits_0^{2\pi} \mathcal{P}^\mu_\pm\of{\sigma} A_\mu }
  \nonumber\\
  &\defas&
  \sum\limits_{n=0}^\infty
  Z^{\mu_1 \cdots \mu_n}\of{\mathcal{P}_\pm}
  \mathrm{Tr}\of{A_{\mu_1}\cdots A_{\mu_n}}
  \nonumber\\
\end{eqnarray}
(where $\mathrm{Tr}$ is the trace and $\mathbf{P}$ indicates path-ordering)
is such a reparameterization invariant functional and in fact all the Taylor
coefficients $Z^{\mu_1 \cdots \mu_n}\of{\mathcal{P}_\pm}$ are, too. These
are the \emph{Pohlmeyer invariants}. They  Poisson-commute with all the
$(\mathcal{P}_\pm)^2\of{\sigma}\,,\forall \sigma$.\\

Using the result \refer{reparameterization property} the representation of the
Pohlmeyer invariants in terms of DDF invariants is immediate, since
\begin{eqnarray}
  \label{Pohlmeyer in terms of DDF}
  Z^{\mu_1 \cdots \mu_n}\of{\mathcal{P}_\pm}
  &=&
  Z^{\mu_1 \cdots \mu_n}\of{\mathcal{P}^R_\pm}
  \,.
\end{eqnarray}
This says that the Pohlmeyer invariants remain unaffected 
under the substitution of ordinary oscillators with DDF invariants 
$\alpha_m^\mu \to A_m^\mu$, $\tilde \alpha_m^\mu \to \tilde A_m^\mu$. 
The level matching condition is fulfilled since all Pohlmeyer invariants are
necessarily of weight 0. \\

The result is therefore that the enveloping algebra of Pohlmeyer invariants is 
a proper subset of the enveloping algebra of DDF invariants. \\

The consistent quantization of the DDF invariants is well known
and yields a quantum algebra which closes on DDF invariants. It therefore
induces a consistent quantization of the algebra of Pohlmeyer invariants
in the sense that the commutator of two quantized Pohlmeyer invariants is
again an invariant (simply because it is again a polynomial in the DDF operators),
though not necessarily a combination of (polynomials of) Pohlmeyer invariants.

This consistent quantization of the \emph{algebra} of the Pohlmeyer inavriants,
just like that of the DDF invariants, exists for every number $D$ of spacetime dimensions.
But, by the well known no-ghost theorem, a representation on a Hilbert space with
positive norm and decoupled longitudinal excitations requires precisely $D=26$
(for the bosonic string).

Since the DDF invariants generalize to the superstring, the prescription
\refer{Pohlmeyer in terms of DDF} also gives us a generalization of the
Pohlmeyer invariants to the superstring.\\

\paragraph{Note:}
After this work was completed we learned of the old articles
\cite{BorodulinIsaev:1982,Isaev:1983} where essentially the same results as given here 
were already reported.
Their relevance for the Pohlmeyer program and for attempts at ``alternative'' 
quantizations of the string seems not to have been widely familiar \cite{Schreiber:2004h}.\\

  I am grateful to Robert Graham for his support and to
  K.-H. Rehren for detailed discussions. I would also like to
  acknowledge interesting discussions with H. Nicolai, B. Schroer and D. Bahns.
  Finally many thanks to A. P. Isaev for making me aware of his work.

This work has been supported by the SFB/TR 12.
\newpage
\bibliography{std}

\end{document}